\documentclass[aps,twocolumn,showpacs]{revtex4}
\usepackage{graphicx}
\usepackage[all]{xy}
\usepackage{amsmath}
\usepackage{amssymb}
\newcommand{\be}{\begin{equation}}
\newcommand{\ee}{\end{equation}}
\newcommand{\ben}{\begin{eqnarray}}
\newcommand{\een}{\end{eqnarray}}
\newcommand{\bes}{\begin{subequations}}
\newcommand{\ees}{\end{subequations}}

\newcommand{\arcsinh}{{\rm arcsinh}}
\newcommand{\sech}{{\rm sech}}
\newcommand{\LL}{{\cal L}}
\newcommand{\LX}{{\cal L}_X}

\newcommand{\LP}{{\cal L}_{\phi}}
\newcommand{\LXX}{{\cal L}_{XX}}

\newcommand{\LXP}{{\cal L}_{X\phi}}

\newcommand{\LPP}{{\cal L}_{\phi\phi}}
\begin{document}

\title{Braneworld Models of Scalar Fields with Generalized Dynamics}
\author{D. Bazeia$^1$, A.R. Gomes$^2$, L. Losano$^1$ and R. Menezes$^1$}
\affiliation{$^1\!$Departamento de F\'{\i}sica, Universidade Federal
da Para\'{\i}ba, 58051-970 Jo\~ao Pessoa, PB, Brazil\\
$^2\!$Departamento de Ci\^encias Exatas, Centro Federal de Educa\c c\~ao Tecnol\'ogica do
Maranh\~ao, 65030-000 S\~ao Lu\'{\i}s, MA, Brazil}
\date{\today}
\begin{abstract}
This work deals with braneworld models driven by real scalar fields with nonstandard dynamics. We develop the first-order formalism for models with standard gravity but with the scalar fields having generalized dynamics. We illustrate the results with examples of current interest, and we find analytical and numerical solutions for warp factors and scalar fields. The results indicate that the generalized braneworld scenario is classically stable, and capable of localizing gravity. 
\end{abstract}
\maketitle

\section{Introduction}

The appearance of electromagnetism in the nineteenth century has triggered a fundamental question, which culminated with the establishment of Special Relativity. The relativity concept is based on the Lorentz group, which was born confronting Galilei invariance and its wrong concept that information may travel at arbitrarily large speed. Not too much later, Special Relativity gave birth to General Relativity, and nowadays, almost one century later, one is facing another consistency problem, this time related to the Planck length.  

Since Relativity makes the speed of light a fundamental constant of nature, it necessarily requires that both space and time change when the coordinate system is changed to another one, with different speed, and this seems to confront with the concept of a fundamental length, the Planck length. We can add to this problem the apparent inconsistency between General Relativity and Quantum Mechanics, and the so-called dark energy problem, which has appeared due to the recent experimental observation that the universe is undergoing accelerated expansion.

The above issues have currently led researchers to consider the possibility of including modifications of the standard scenario involving matter and geometry. Several possible ways are under consideration, and a very popular procedure is known as the quintessence way, in which one in general includes dynamical scalar fields that can interact through a diversity of possibilities. 

In the present work, we will focus our attention on the braneworld scenario, but we will deal with scalar fields with nonstandard kinetic terms coupled with standard gravity. In the well-known Randall-Sundrum model \cite{RS}, we can further add scalar fields \cite{GW} with usual dynamics and allow them to interact with gravity in the standard way. In this scenario, the smooth character of the solutions generate thick brane with a diversity of structures \cite{first1,first2,OT}. However, even when one look for static solutions, the intrinsic nonlinear character of the Einstein equations usually result in an intricate system of coupled ordinary differential equations that are hard to solve. Despite the possible numerical treatment, it is also of interest to find models which support analytic solutions. In this particular, one can consider specific situations where first-order differential equations appear describing the scalar field and metric functions, with the potential having a very specific form \cite{first1,first2,OT}.

In recent years, there appeared some interesting models with non canonical dynamics with focus on early time inflation or dark energy, as good candidates to solve the coincidence problem \cite{AP}. These kind of models have also been discussed in investigations of topological defects. In fact, global topological defects have been considered in \cite{Babichev:2006cy,Bazeia:2007df,Adam:2007ij,Babichev:2007tn}. For instance, in Ref.~\cite{Babichev:2006cy} one has found domain walls, global strings and global monopoles, and in Ref.~\cite{Bazeia:2007df} some formal aspects of unidimensional topological solutions has been studied, and there it was shown that the linear stability is preserved for some classes of models. Also, in  \cite{Adam:2007ij} it was shown that quartic potential can support compacton solutions for a specific Lagrangian, and some local vortices were investigated in \cite{Babichev:2007tn}. In Ref.~\cite{Olechowski:2008bh}, the generalized models are used for stabilization of inter-brane distance. The brane study was also considered in two other works \cite{Adam}, under the action of specific scalar field model which gives rise to compacton solutions. 

The search for analytical solutions for such generalized models is a non trivial task. Our motivation has arisen from a previous investigation, in which one considers first-order differential equations to solve the corresponding equations of motion \cite{blm}. The presence of first order equations simplifies the investigation and can yield to other extensions, as we will show in this work, where we modify the standard braneworld scenario with the inclusion of scalar fields with nonstandard dynamics.

We develop the investigations as follows: in the next Sec.~\ref{model} we study $(4,1)$ brane models with nonstandard kinetic term coupled with gravity. As usual, we suppose that both the scalar field and the warp factor depend only on the fifth extra dimension. After an Ansatz for the metric characterizing $M_4$ branes with an asymptotically $AdS_5$ bulk is implemented, the general structure of the equations of motion are obtained. We show from stability analysis that metric and scalar field perturbations can be decoupled in the transverse traceless gauge and that the spectrum of excitations produces a zero-mode and a continuum of positive massive modes. This shows that the proposed generalized scenario is stable and capable of localizing gravity in a way similar to the standard case. Encouraged by this general result, in the next Sec.~\ref{brane} we implement the first-order framework put forward in \cite{blm}, in order to investigate two distinct families of models. In Sec.~\ref{end} we end the work with some comments and conclusions. 

The generalized model that we consider has the form ${\LL}= F(X)-V(\phi)$, and below we study two specific forms for the nonstandard kinetic term. The first one is given by $F=X+\alpha |X|X$ and depends on a parameter $\alpha$ which drives the model away from the standard case. Analytical expressions for small $\alpha$ are obtained and compared with the numerical investigation done for a larger range of values of $\alpha$. We investigate the energy density of the brane as well as the necessary conditions for gravity localization and modifications of Newton's law, and there we show that the numerical study confirms the analytical results obtained for small $\alpha.$ Here an interesting result is that gravity localization seems to be more effective at smaller values of $\alpha$,
showing that the robustness of the model seems to weaken for increasing values of $\alpha$. The second model is described by $F=-X^2$, and although it is well distinct from the former one, our classical investigation show that it is also capable of localizing gravity.  

\section{Generalized dynamics}
\label{model}

We start with a five-dimensional action in which gravity is coupled to the scalar field in the form
\ben\label{action1}
S=\int d^5 x \sqrt{g} \left(-\frac14 R +{\cal L}(\phi, X) \right)
\een
where we are using $G^{(5)}=1/(4\pi)$ and the signature of bulk metric as $(+----)$, with $g=\det(g_{MN})$. We take the spacetime coordinates and fields as dimensionless quantities,  and the convention $M,N=0,1,2,3,4$ and $\mu,\nu=0,1,2,3$. We also define the invariant 
\ben
X=\frac12 \nabla^M\phi \nabla_{\!M}\phi
\een
The Einstein equations are $G_{AB}=2\,T_{AB}$, with the energy-momentum tensor having the form 
\ben
T_{AB}=\nabla_A \phi \nabla_B \phi\, \LX - g_{AB} \LL
\een
The equation of motion for the scalar field is given by
\ben
G^{AB}\nabla_A \nabla_B \phi + 2 X {\cal L}_{X\phi} - {\cal L}_\phi = 0
\een
where $G^{AB}$ has the form
\ben
G^{AB}={\cal L}_X g^{AB} + {\cal L}_{XX} \nabla^A \phi \nabla^B \phi
\een
We use the notation $\LX=\partial \LL/\partial X$ and $\LP=\partial \LL / \partial \phi$, etc. In order for the above differential equation to be hyperbolic, the condition 
\ben
\frac{\LX+2X\LXX}{\LX} > 0
\een
must be fulfilled.

We use the standard notation, and write the metric as
\ben\label{metric}
ds^2=e^{2A}\eta_{\mu\nu}dx^\mu  dx^\nu  - dy^2
\een
where $A=A(y)$ describes the warp factor and only depends on the extra dimension $y$. As usual, we suppose that the field $\phi$ is static, and also, it only depends on the extra dimension. Thus, $A=A(y)$ and $\phi=\phi(y),$ and so the equation of motion for the scalar field reduces to
\ben\label{SEM}
\label{eqs}
\left(\LL_X+2X\LL_{XX}\right) \phi^{\prime\prime}-(2X\LL_{X\phi} - \LP)=-4 \LL_X \phi^\prime A^\prime \label{motioneq1}
\een
where prime denotes derivative with respect to the extra dimension. The Einstein's equations with the metric (\ref{metric}) lead to
\bes\label{EE}\ben
A^{\prime\prime}&=&\frac{4}{3}X \LL_X  \label{motioneq2}\\
A^{\prime 2} &=& \frac13 \left(\LL - 2X\LL_X \right) \label{motioneq3}
\een\ees
where for static solutions we have $X=-\phi^{\prime2}/2$.  The equations \eqref{SEM} and \eqref{EE} are not independent, the last one being the null energy condition that imposes a brane with positive pressure, obeying ($\LL - 2X\LL_X>0$). In particular, we can multiply (\ref{motioneq1}) by $\phi'$ in order to get
\ben
(\LL -2 X\LX )^{\prime}=-4 \phi^{\prime2} A^\prime \LX \label{eqQuadra}
\een
and now, if we substitute Eq. (\ref{motioneq3}) we then recover Eq. (\ref{motioneq2}). 

We note that the Eqs.~(\ref{eqs})-(\ref{motioneq3}) reduce to the known equations in the standard case, in which $\LL=X-V:$
\bes\ben
&&\phi^{\prime\prime}+4  \phi^\prime A^\prime+ V_\phi=0 \label{motioneq12}
\\
&&A^{\prime\prime}+\frac23 \phi^{\prime2}=0 \label{motioneq22}
\\
&&A^{\prime 2}-\frac{1}{6}\phi^{\prime2}+\frac13 V(\phi)=0 \label{motioneq31}
\een\ees

An important characteristic of the brane is its tension, which is given by
\ben\label{Tensioneq}
{\cal T} = \int dy e^{2A(y)}T_{00} = \int dy \rho,
\een
where $\rho(y) = -e^{2A(y)} {\cal L}$ is the energy density.

The proposed investigation may be of direct interest to high energy physics, but it is important to know if the modification of the scalar field dynamics will contribute to destabilize the geometric degrees of freedom of the braneworld model. We investigate this issue studying linear stability in the usual way. We consider metric perturbations in the form  
\ben
ds^2= e^{2A(y)}(\eta_{\mu\nu}+h_{\mu\nu}(y,x)) dx^{\mu}dx^\nu - dy^2
\een
We must also consider fluctuations of the scalar field
\ben
\phi=\phi(y)+\tilde \phi(y,x)
\een
The first order contribution of the fluctuations to the scalar $X$ is written as $\tilde X^{(1)}=(1/2) h^{\mu\nu} \partial_\mu \phi \partial_\nu \phi + \partial^\mu \phi \partial_\mu \tilde\phi.$ We found the first order contributions of Einstein equations in Ricci form as $R_{AB}=\bar{T}_{AB}$, with $\bar{T}_{AB}=T_{AB}$ - $(1/3) g_{AB} T^C_{\mbox{ }C}$, and
\bes\ben
{\bar T }_{\mu\nu}^{(1)}&=&\frac23 \eta^{\mu\nu} e^{2A}\nonumber
\\
&&\left(-X\left(\LXP \tilde \phi - \LXX \phi^\prime \tilde\phi^\prime\right)+\LP \tilde \phi\right)\nonumber
\\
&& -2e^{2A}h_{\mu\nu}\left(X\LX -\LL\right)
\\
{\bar T }_{\mu4}^{(1)}&=&\LX \phi^\prime \nabla_\mu \tilde \phi \\
{\bar T }_{44}^{(1)}&=&-\frac23 \left(2\LXP X +\LP\right)\tilde \phi\nonumber
\\
&&+\frac23 \left(2\LXX X + 3 \LX\right)\phi^\prime {\tilde\phi}^\prime
\een\ees
In this case, Einstein's equations turn out to be
\ben
 e^{2A}\left(1/2 \partial^2_y + 2 A^\prime \partial_y\right) h_{\mu\nu} +\frac12 \eta_{\mu\nu} e^{2A} a^\prime \partial_y (\eta^{\alpha\beta}h_{\alpha\beta}) \nonumber
\een
\ben 
-\frac12 \eta^{\alpha\beta} \left(\partial_\mu \partial_\nu h_{\alpha\beta}-\partial_\mu\partial_\alpha h_{\nu \beta} -\partial_\nu\partial_\alpha h_{\mu \beta}\right)\nonumber
\een
\ben
=\frac43 e^{2A}\eta_{\mu\nu} \left(-X\left(\LXP \tilde \phi - \LXX \phi^\prime \tilde\phi^\prime\right)+2\LP\tilde\phi\right)\label{eqperturb1}
\een
and
\ben
&&\frac12 \eta^{\alpha} \partial \left(\partial_\alpha h_{\mu\beta}-\partial_\mu h_{\alpha\beta}\right)=\LX \phi \partial_\mu \tilde \phi
\nonumber
\\
&&-\frac12 \left(\partial^2_y + 2 A^{\prime2}\partial_y \right)\eta^{\alpha\beta}h_{\alpha\beta}  \nonumber
\\
&&\!=\!-\frac23 \left(2\LXP X\!+\!\LP\right)\tilde \phi\!+\!\frac23 \left(2\LXX X\!+\!3\LX\right)\phi^\prime {\tilde\phi}^\prime
\een
The equation of motion for the scalar field gives
\ben
&&\LX e^{2A}\Box\tilde\phi\!-\!\left((2\LXX X\!\!+\!\LX)\tilde\phi^\prime \right)^\prime
\!\!\!-\!4 A^\prime(2\LXX X\!\!+\!\LX)\tilde \phi^\prime \nonumber
\\
&&-\left(4 \LXP \phi^\prime A^\prime\!+\! \left(\LXP \phi^\prime \right)^\prime\! +\! \LPP\right)\tilde \phi = 
\LX \phi^\prime \eta^{\alpha\beta}h_{\alpha\beta}^\prime
\een

Let us now consider the transverse traceless components for metric fluctuations
\ben
\bar{h}_{\mu\nu}=\left(\frac{1}{2}(\pi_{\mu\alpha}\pi_{\nu\beta}+\pi_{\mu\beta}\pi_{\mu\alpha})-\frac13 \pi_{\mu\mu}\pi_{\alpha\beta}\right)h^{\alpha\beta}
\een
where $\pi_{\mu\nu}=\eta_{\mu\nu} -\partial_\mu \partial_\nu/\Box.$ We note that the net effect of this projection operation is to decouple the metric fluctuation equation from the scalar field equation, even in the general case which is being considered in the present work. Indeed, we can check that equation (\ref{eqperturb1}) reduces to the known equation
\ben
\left(\partial^2_y + 4 A^\prime \partial_y - e^{-2A}\Box\right)\bar{h}_{\mu\nu}=0
\een

The next steps are known: we introduce the $z$-coordinate in order to make the metric conformally flat, with $dz=e^{-A(y)}dy$ and we write
\ben
H_{\mu\nu}(z)=e^{-ipx}e^{3/2A(z)}\bar{h}_{\mu\nu}
\een
In this case, the 4-dimensional components of ${\bar h}_{\mu\nu}$ obey the Klein-Gordon equation and the metric fluctuations of the brane solution lead to Schr\"odinger-like equation
\ben
\left[-\partial_z^2 + U(z) \right]H_{\mu\nu} = p^2 H_{\mu\nu}
\een
where 
\ben
U(z)=\frac{9}{4}A^{\prime2}(z) + \frac32 A^{\prime\prime}(z)
\een
We can write this equation in the form
\ben
\left(\partial_z + \frac32 A^{\prime} (z)\right)\left(-\partial_z + \frac32 A^{\prime}(z) \right) H_{\mu\nu} = p^2 H_{\mu\nu}
\een
This factorization directly shows that there are no graviton bound-states with negative mass, and the graviton zero mode $H_{\mu\nu}(z) \propto e^{\frac{3}{2}A(z)}$ is the ground-state of the associated quantum mechanical problem.

This result leads to the important conclusion that the modification of the scalar field dynamics does not contribute to destabilize the geometric degrees of freedom which appears in the standard braneworld scenario. Thus, the modification here proposed is robust and may be of direct interest to high energy physics.

\section{The braneworld scenario}
\label{brane}

Let us start reviewing the case without gravity, setting $A(y)=0$, which means that only the scalar field equation of motion has to be considered. 
We follow \cite{Bazeia:2007df,blm} and we get 
\ben
\left(\LX+2X\LXX\right) \phi^{\prime\prime}=2X\LXP - \LP
\een
In this case, we can use Eq. (\ref{eqQuadra}) to get $\LL - 2 \LX X = C$, where $C$ is an integration constant which can be identified with the pressure $T^{44}$ in the absence of gravity. For stable configurations, the pressureless condition is necessary. Thus we write
\ben\label{Zeroflatoeq}
\LL - 2 \LX X =0	
\een
This equation depends on the scalar field and its first derivative. Therefore, it is a first-order equation. The tension of the solution is
\ben
{\cal T}=-\int^{\infty}_{-\infty} dy \LL = \int^{\infty}_{-\infty} dy \LX \phi^{\prime2}
\een
If we introduce the function $W=W(\phi)$ such that
\ben
\LX \phi^\prime=W_\phi
\een
we can write the tension in the form
\ben
{\cal T}=W(\phi(\infty)) - W(\phi(-\infty))
\een
The interesting thing here is that the tension does not depend on the explicit form of the solution, but only on its asymptotic values. 

Let us now consider some explicit examples. For instance, we can deal with 
\ben\label{type1}
\LL=X+\alpha |X|X - V(\phi)
\een
where $\alpha$ is a real, non negative parameter. We name this the type I model. Alternatively, we could choose the function
\ben
\LL=X-\alpha X^2 - V(\phi) 
\een
but this does not change the classical scenario, as highlighted in Ref.\cite{Adam}. Of course, if $\alpha=0$ the standard scenario is restaured. For $\alpha$ a very small parameter, we can investigate the contribution of this term as a perturbation to the standard scenario. This leads us with the expressions 
\ben
\phi^\prime &=& W_\phi - \alpha\, W_\phi^3 
\\
V(\phi)&=&\frac12 W_\phi^2 - \frac{\alpha}{4}W_\phi^4
\een
We can also consider another model, for instance 
\ben\label{type2}
{\cal L}= -X^2 - V(\phi)
\een
We name this the type II model, and here the first order formalism leads to the equations
\ben
\phi^\prime &=& W_\phi^{\frac13} \\
V(\phi)&=&\frac34 W_\phi^{\frac43}
\een
More details of the first-order formalism in flat spacetime can be seen in \cite{Bazeia:2007df,blm}. 

To extend the first-order framework to the braneworld scenario, we follow some works in Ref.~\cite{first2} and choose the derivative of the warp factor with respect to the extra dimension to be a function of the scalar field, and we write
\ben\label{firstorder1}
A^\prime=-\frac13 W(\phi)
\een
This equation also appear in the standard braneworld scenario. Our point here is that since in the action (\ref{action1}) the geometric sector remains unchanged, we expect that this equation remains unchanged too. In this case, we use the equation (\ref{motioneq2}) to write 
\ben\label{firstorder2}
\LX \phi^\prime = \frac{1}{2} W_\phi
\een
which is the same that appears in the absence of gravity.  The null energy condition (\ref{motioneq3}) can be written as
\ben\label{firstorder3}
\LL - 2 \LX X =\frac13 W^2 
\een
The equations (\ref{firstorder2}) and (\ref{firstorder3}) impose a constraint in the Lagrange density. It is not difficult to show that these first-order equations solve the second-order equation of motion (\ref{eqs}).
 
Let us consider the standard braneworld model.  The Lagrange density for the scalar field is given by
\ben
{\LL}= X - V(\phi)
\een
The set of equations (\ref{firstorder1}), (\ref{firstorder2}) and (\ref{firstorder3}) give
\bes
\ben
\phi^\prime &=&  \frac12 W_\phi
\een
and the constrained potential is
\ben\label{standarV}
V(\phi)=\frac18 W_\phi^2 - \frac13 W^2
\een
\ees
These equations describe BPS solutions since they appear in supergravity \cite{first1}. 

Now we consider the case in which the scalar field has generalized dynamics. The general structure of the Lagrange density is given by
\ben
\LL= F(X) - V(\phi)
\een
For such models the scalar field is sometimes called a k-field. The equation of motion is
\ben
(F^{\prime} +2X F^{\prime\prime}  ) \phi^{\prime\prime} -  V_\phi = - 4 F^\prime \phi^\prime A^\prime,
\een
and from the Einstein equations we obtain 
\ben
A^{\prime\prime}=\frac{4}{3}  F^\prime X,\;\;\;\;\;
A^{\prime2}=\frac{1}{3} \left(F-V-2XF^\prime \right)
\een
We notice that in the standard situation $F(X)=X,$ the above equations lead to the standard braneworld case. In the generalized situation, the first-order equations are (\ref{firstorder1}) and
\bes
\ben\label{eq111}
F^\prime  \phi^\prime &=& \frac{1}{2} W_\phi \label{FWphi1}\\
F-2F^\prime X- V(\phi)  &=&\frac13 W^2\label{eq3} 
\een
\ees
The equation \eqref{eq111} has the form $G(\phi^\prime) = \frac{1}{2} W_\phi$. For some models this can be rewritten as
\ben
\phi^\prime=G^{-1}\left(\frac{1}{2} W_\phi\right)
\een 
Now, substituting this into Eq.~\eqref{eq3} leads to the potential
\ben\label{Vfirsto111}
V(\phi)=(F-2F^\prime X)|_{\phi^\prime=G^{-1}(\frac{1}{2} W_\phi)}-\frac13 W^2 
\een
Comparing this expression with Eq. (\ref{standarV}) for the potential of the standard case, we see that only the $W_\phi$ portion is changed. The reason for this is that the $W$ portion follows from a pure geometric contribution which remained unchanged. 

We now illustrate the investigations with the type I and type II models, as described in \eqref{type1} and \eqref{type2}, respectively.

\subsection{Type I model}
\label{sec:typeI}

We use the function $F=X+\alpha |X|X$. In this case, the equation of motion for the scalar field becomes
\ben
\phi^{\prime\prime}+4\phi^\prime A^\prime -  V_\phi =\alpha(3 \phi^{\prime\prime} - 4 \phi^{\prime}  A^\prime)\phi^{\prime2},
\een
Using the first-order equations, we see that the scalar field has to obey 
\ben\label{firstF1}
\phi^\prime + \alpha \phi^{\prime3} = \frac12 W_\phi
\een
This algebraic equation of third degree for $\phi^\prime$ has the only real solution
\ben
\phi^\prime &=&\frac{m(W_\phi)}{6\alpha} -\frac{2}{m(W_\phi)} \label{egiireerpo}
\een
where
\ben
m(W_\phi)=(54\alpha^2\,W_\phi\! +6\sqrt{3}(16\,\alpha^3\! + 27 \alpha^4\,W_\phi^2)^{1/2})^{1/3}
\een
From Eq. (\ref{Vfirsto111}) we can write the potential as 
\ben
V(\phi)=\frac12\phi^{\prime2}+ \frac34\alpha\phi^{\prime4} -\frac13 W^2
\een
or then, substituting (\ref{egiireerpo})
\ben\label{potgen}
V(\phi)&=&\frac12 \left(\frac{m(W_\phi)}{6\alpha}-\frac{2}{m(W_\phi)}\right)^2\nonumber
\\
&&+\frac{3\alpha}{4}\left(\frac{m(W_\phi)}{6\alpha}-\frac{2}{m(W_\phi)}\right)^4\!\!-\frac13 W(\phi)^2
\een

In order to ease investigations, let us focus our study in the case of $\alpha$ very small. Here we get, up to first-order in $\alpha,$ the field equation
\ben\label{eq2brane222}
\phi^\prime=\frac12 W_\phi - \frac{\alpha}{8} W_\phi^3
\een
with the corresponding potential 
\ben
V(\phi)=\frac{1}{8}W_\phi^2 - \frac{\alpha}{64} W_\phi^4 - \frac13 W^2 
\een
The equation (\ref{eq2brane222}) yields, after an integration, 
\ben
2\int\frac{d\phi}{W_\phi}+\frac{\alpha}{2} W(\phi)=y
\een
and so $\phi(y) = \phi_0\left(y-\alpha W(\phi_0)/2\right)$, 
where $\phi_0(y)$ is the solution when $\alpha$ vanishes. We expand this solution to get
\ben
\phi(y) = \phi_0(y) - \frac\alpha2 \,\phi^\prime_0(y)  W(\phi_0(y))
\een
or using (\ref{eq2brane222})
\ben\label{relationphi}
\phi(y) = \phi_0(y) - \frac\alpha4  W_\phi(\phi_0(y))	  W(\phi_0(y))
\een

The warp factor obeys the equation
\ben
A^\prime = -(1/3) W(\phi_0(y) - (\alpha/2) \,\phi^\prime_0(y)  W(\phi_0(y)))
\een
It is then easy to see that
\ben\label{warpalpha}
A(y) = A_0(y) + \frac\alpha{12} W(\phi_0(y))^2 
\een
where $A_0$ represents the standard warp factor, for $\alpha=0$. The brane tension (\ref{Tensioneq}) is
\ben
{\cal T}=\int dy e^{2A(y)}\left(\frac14 W_\phi^2 - \frac13 W^2 - \alpha \frac{W^4_\phi}{16}\right)
\een
or better
\ben
\!{\cal T}\!\!=\!{\cal T}_0\!-\!\frac\alpha{48}\!\!\int\!\!dy e^{2A(y)}\!\!\left(6 WW_\phi^3\! -\! 8 W^2W_\phi^2\! +\! 3 W_\phi^4 \right)\!|_{\phi=\phi_0}
\een
where ${\cal T}_0$ is the tension of the brane in the standard scenario. 

We can consider the explicit example, with $W(\phi)$ given by
\ben
W(\phi)=3a\sin(b \phi), 
\een

\begin{figure}[htbp]
\centering
\includegraphics[width=6.0cm,height=5cm]{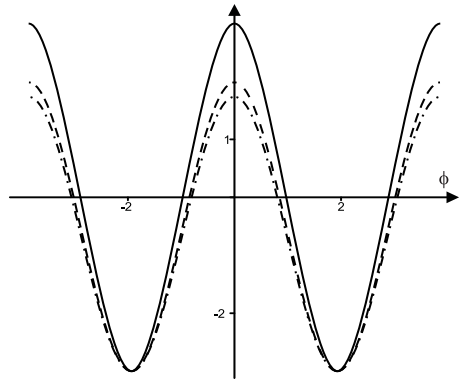}
\caption{Plots of the potential \eqref{potgen} for $b=\sqrt{6}/3$, $a=1$ and for $\alpha=0.1$ (solid line), $\alpha=1$ (dashed line) and $\alpha=10$ (dot-dashed line).}
\end{figure}

In this case the potential for $\alpha$ small has the form 
\begin{eqnarray}\label{poten}
V\!=\!\frac98 a^2 b^2\! \cos^2(b\phi)\!-\!3a^2\! \sin^2(b\phi)\!-\!\frac{81\alpha}{64}a^4b^4\!\cos^4(b\phi)
\end{eqnarray}
In Fig.~1, we plot the potential for some values of $\alpha,$ not necessarily small. We have also depicted the other Figs.~2, 3, and 4 below, numerically, for several values of $\alpha$, and we have also checked that the corresponding analytic expressions, obtained for $\alpha$ very small, completely agrees with the numerical results with $\alpha$ small. We then note that the numerical study give full support to the analytical expressions which we have obtained for $\alpha$ very small. We further note that the numerical study shows the robustness of the model for a large range of possibilities for the parameter $\alpha.$ 

We can use (\ref{relationphi}) to write for the scalar field $\phi(y)=\phi_0(y)-(9\alpha a^2 b/8)\sin(2b\phi_0)$, or explicitly 
\begin{eqnarray}
\phi(y)&=&\frac1b \arcsin\left[\tanh\left(\frac32 a b^2 y\right)\right] \nonumber \\
&& - \frac{9\alpha a^2b}4 \tanh\left(\frac{3}{2} a b^2 y\right) \sech\left(\frac32 a b^2 y\right)
\end{eqnarray}
Fig.~2 shows the kink profile (upper panel), with the $\alpha$ parameter increasing the brane thickness, as can also be seen from the plots of the energy density (lower panel). This is an interesting result, since it shows that the $\alpha$-dependent term used to modify the dynamics of the scalar field contributes to thicker the brane.   
\begin{figure}[th]
\begin{center}
\includegraphics[{width=6.0cm},height=5cm]{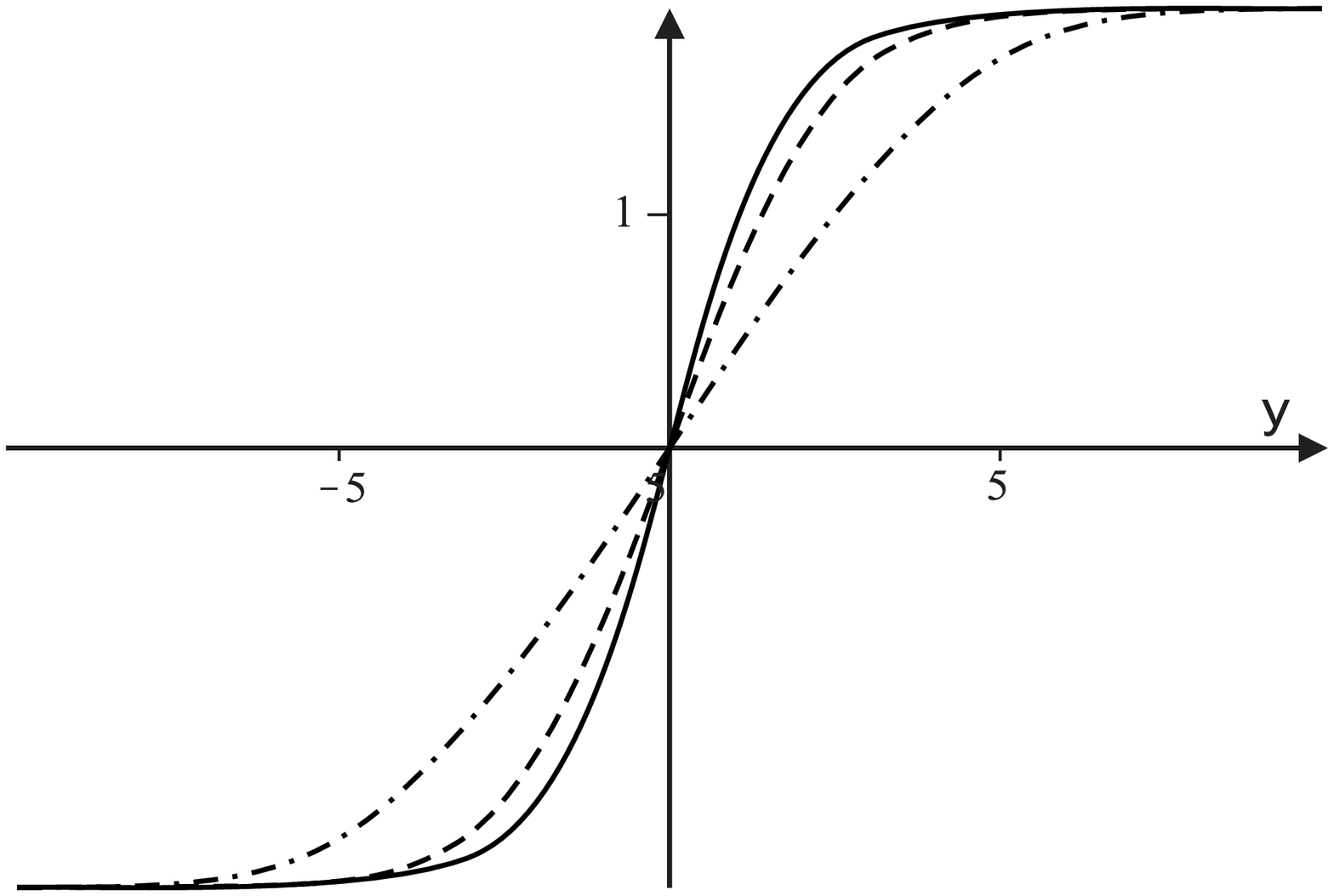}
\includegraphics[{width=6.0cm},height=5cm]{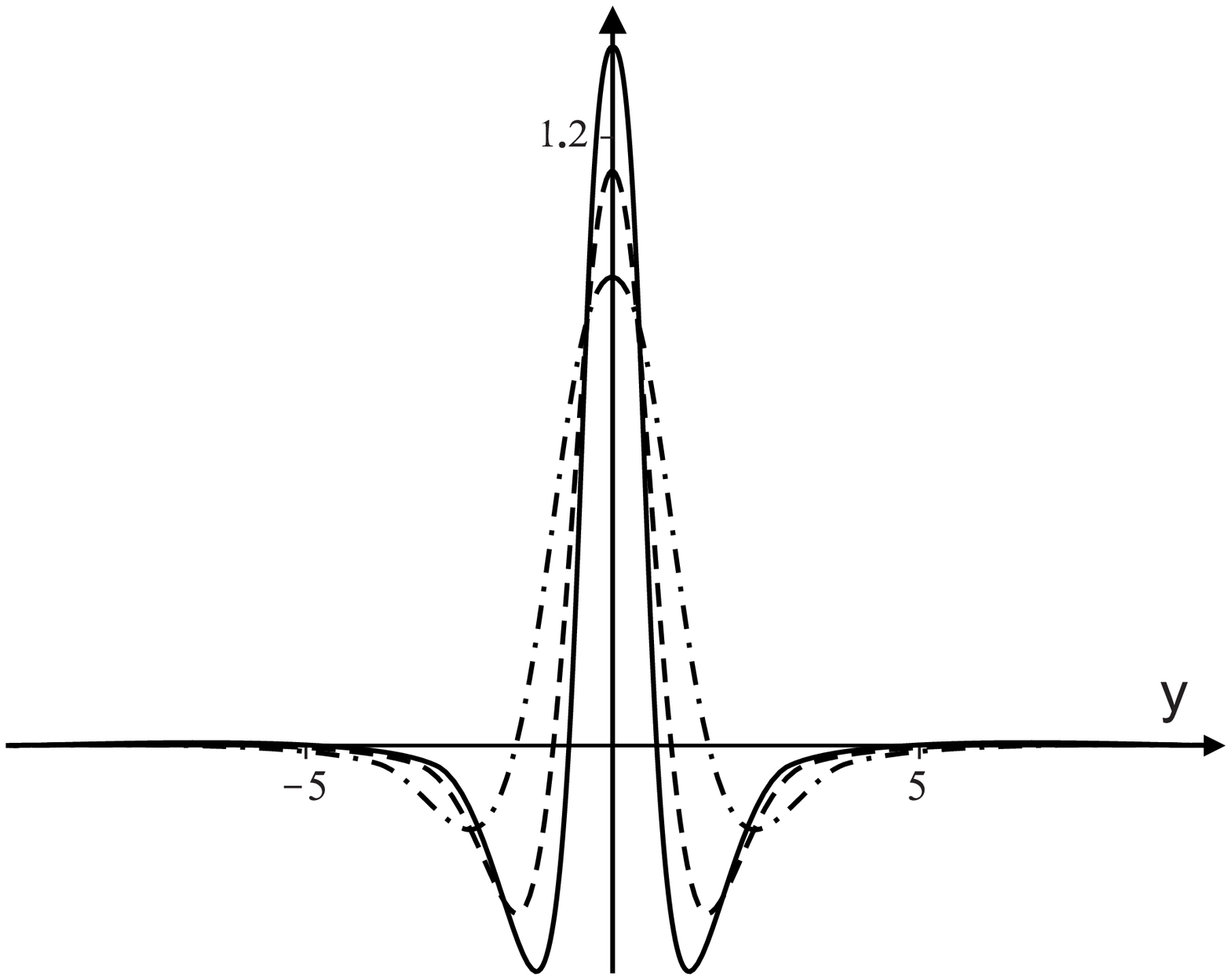}
\end{center}
\caption{Plots of the scalar field $\phi(y)$ (upper panel) and the energy density $\rho(y)$ (lower panel) for $\alpha=0.1$ (solid line), $\alpha=1$ (dashed line) and $\alpha=10$ (dot-dashed line). We use $b^2=2/3$.}
\end{figure}

We use Eq.~(\ref{warpalpha}) to get
\begin{eqnarray}
A(y)\!=\!\frac{2}{3b^2}\!\ln\!\left(\sech\left(\frac32ab^2 y\right)\!\right)\!\!+\!\!\frac{3a^2\alpha}{4}\!\tanh^2\left(\frac{3}{2}ab^2 y\!\right)
\end{eqnarray}
which is depicted in Fig.~3.
\begin{figure}[th]
\begin{center}
\includegraphics[{width=6.0cm},height=5cm]{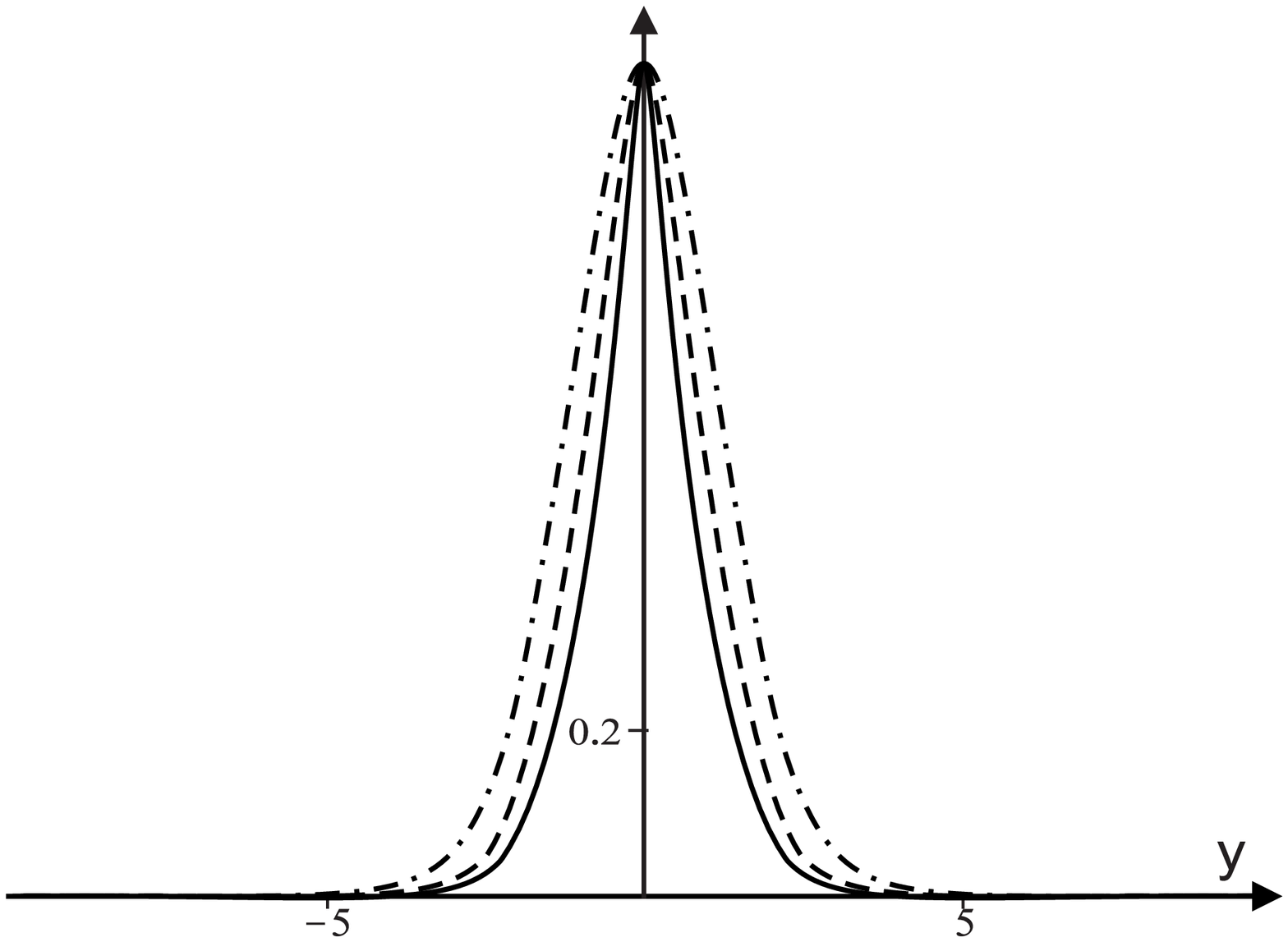}
\includegraphics[{width=6.0cm},height=5cm]{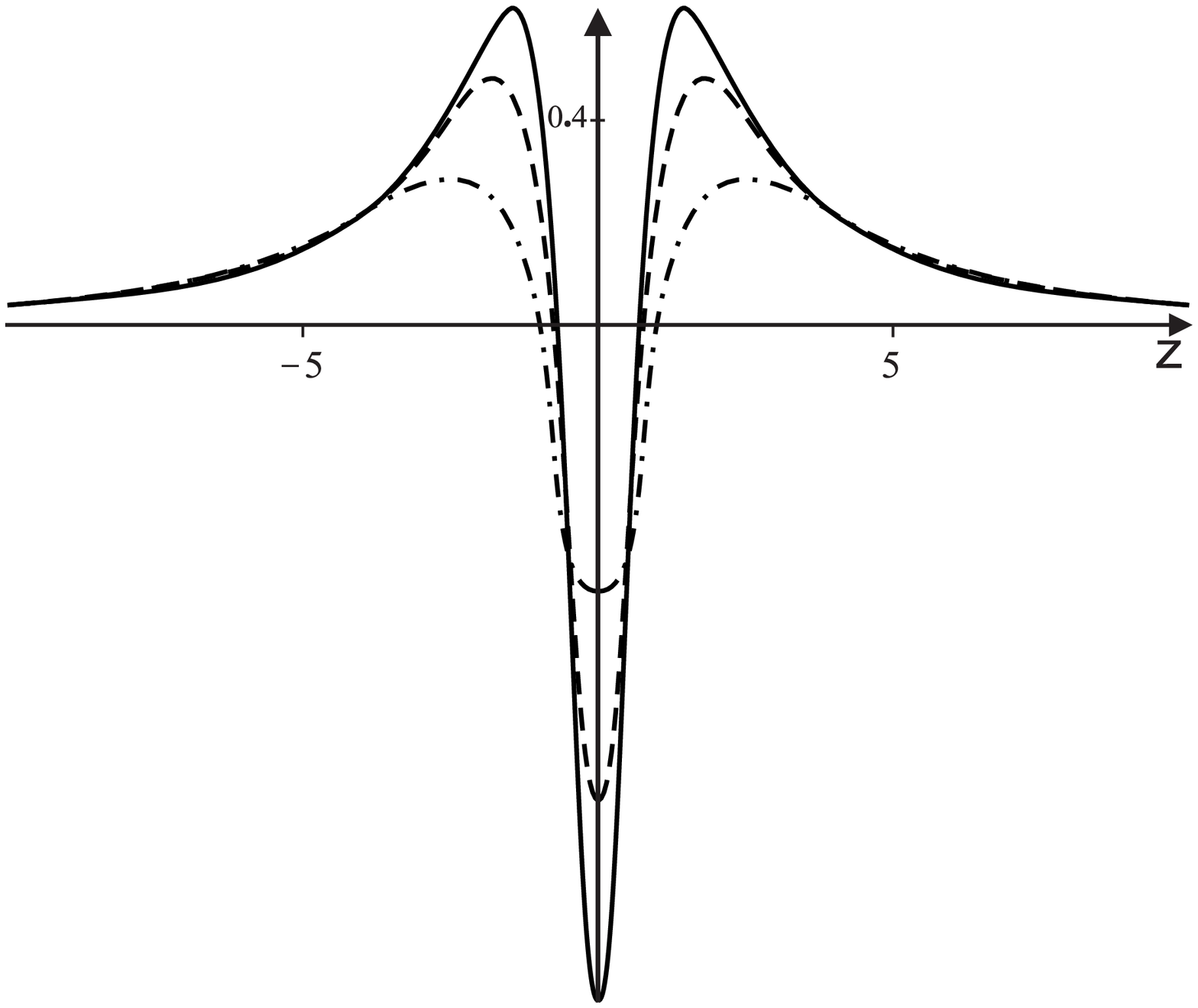}
\end{center}
\caption{Plots of the warp factor $e^{2A(y)}$ (upper panel) and the Schr\"odinger-like potential $U(z)$ (lower panel) for $\alpha=0.1$ (solid line), $\alpha=1$ (dashed line) and $\alpha=10$ (dot-dashed line). We use $b^2=2/3$.}
\end{figure}
For $b^2=2/3$, we use the transformation $dz=e^{-A(y)}dy$ to write
\ben
dz=\cosh(ay)-\frac{3a^2}{4}\sinh^2(ay)\sech(ay)
\een
where
\ben
z=\frac{\sinh(ay)}{a}+\frac{3a\alpha}{4}(\arctan(\sinh(ay))-\sinh(ay))
\een
\ben
y=\frac1a \arcsinh(az) - \frac{3a\alpha}{4} \frac{(\arctan(az)-az)}{\sqrt{1+a^2z^2}}
\een
The warp factor is now written in terms of the $z$ variable
\ben
\label{typeIAz}
A(z)=-\frac12 \ln(1+a^2 z^2) + \frac{3a^3\alpha}{4} \frac{z\,\arctan(az)}{{1+a^2z^2}}
\een
The Schr\"odinger-like potential is given by
\ben
U(z)=\frac{9}{4}A^{\prime2}(z) + \frac32 A^{\prime\prime}(z)
\een
and has the explicit form
\begin{eqnarray}
U(z)\!&=&\!\frac{3a^2(5a^2z^2-2)}{4(1+a^2z^2)^2}+\!\frac{9}{8} \frac{a^4\alpha}{{ \left( 1+{a}^{2}{z}^{2} \right) ^{3}}}\nonumber\\
&& \left({{{a}z( 5{a}^{2}{z}^{2}-9) \arctan( az)}}\!+\!{(2-\!7\,{a}^{2}{z}^{2})} \right)
\end{eqnarray}

In Fig.~3 (lower panel) we plot the potential $U(z)$ for several values of $\alpha$. We see that the characteristic volcano profile for dynamically generated $M_4$ brane immersed in an asymptotically $AdS_5$ bulk. The increase of $\alpha$ leads to a reduction of the maximum of the potential, and this may modify the way gravity is localized in the brane. Thus, we have to investigate if this behavior produces any sensible effects on the localization of gravity in the brane
generated in this case. 

\begin{figure}[th]
\begin{center}
\includegraphics[{width=6.0cm},height=5cm]{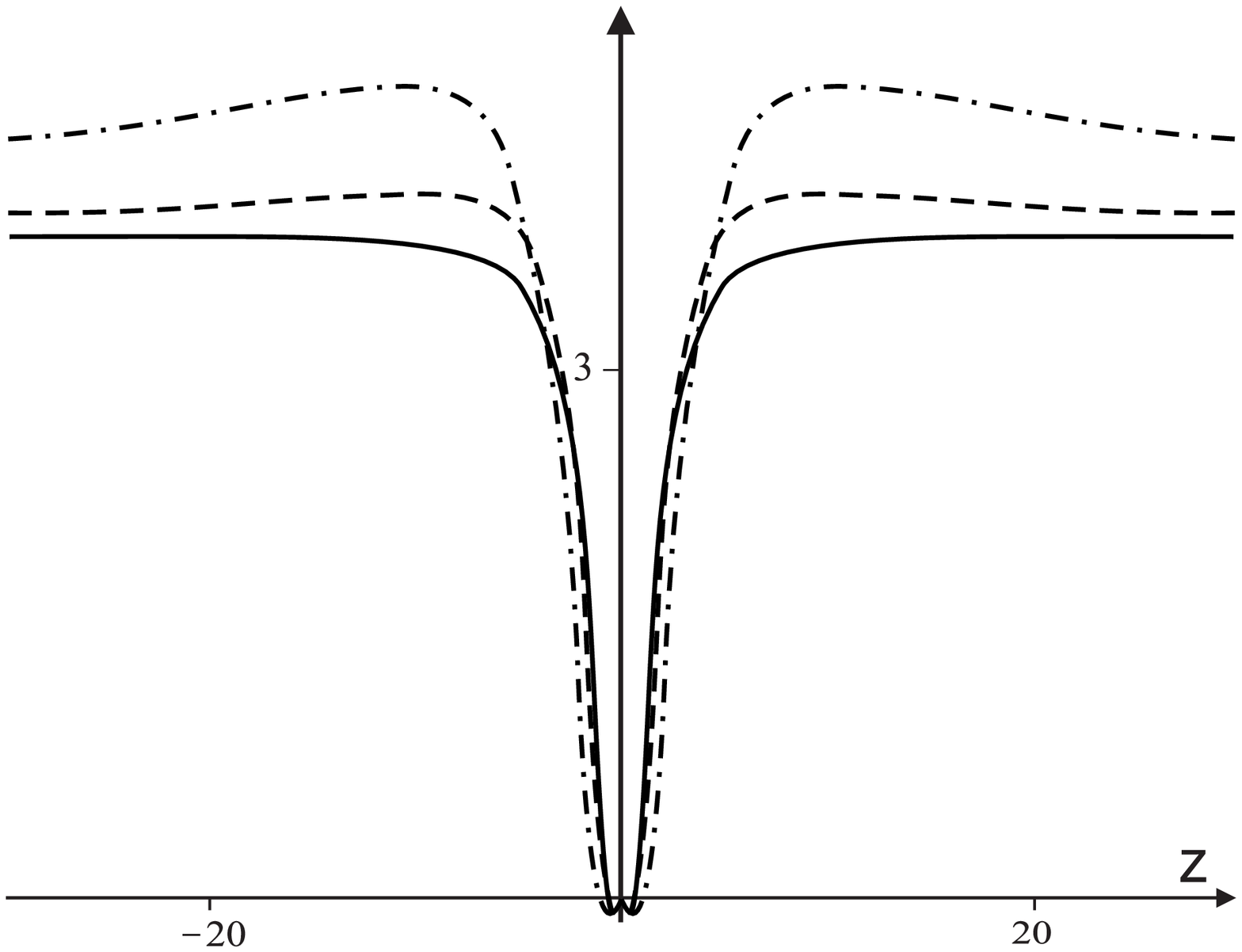}
\includegraphics[{width=6.0cm,height=5cm}]{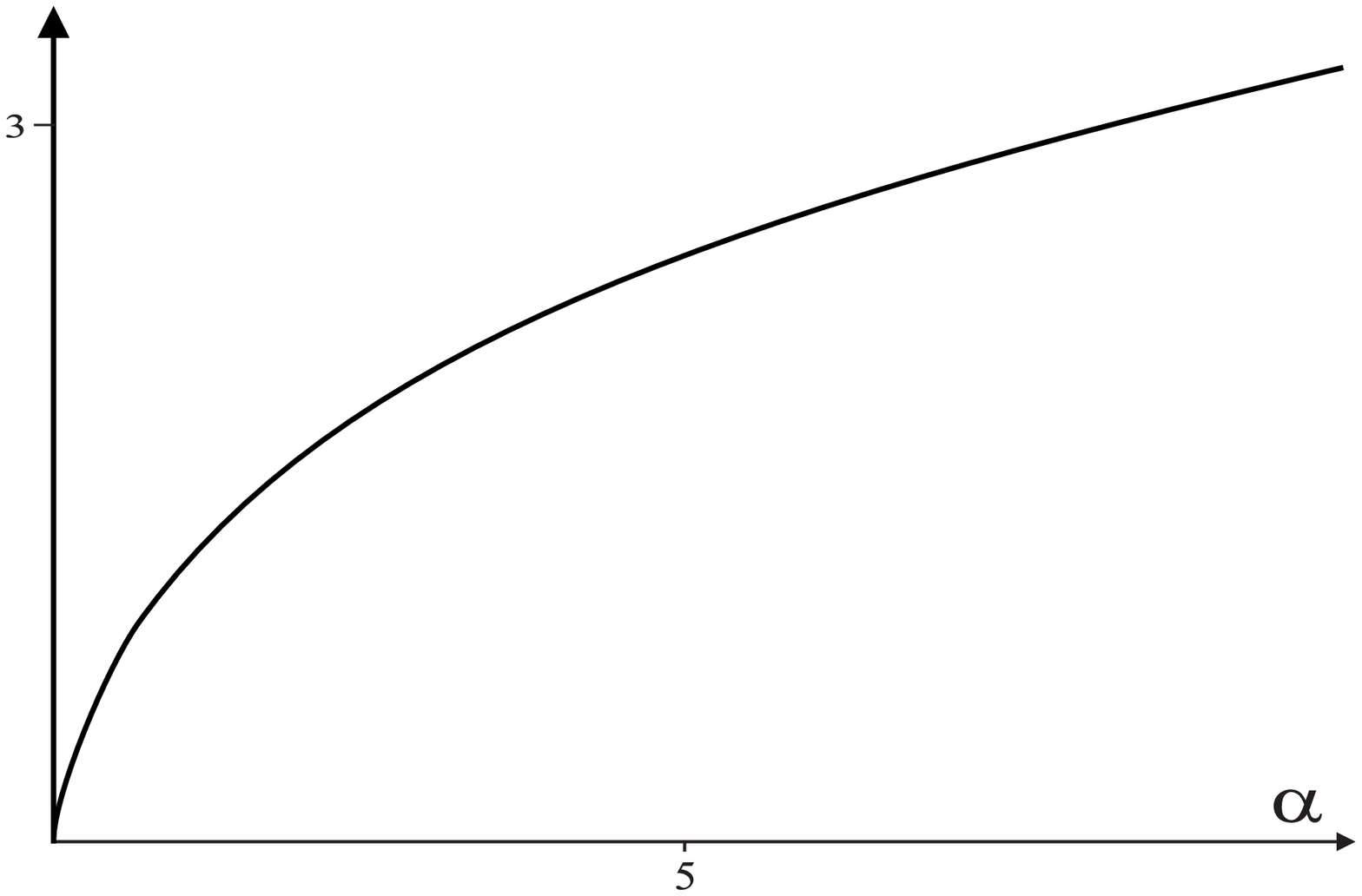}
\end{center}
\caption{Plots of $z^2U(z)$ for the type I model (upper panel) for $\alpha=0.1$ (solid line), $\alpha=1$ (dashed line) and $\alpha=10$ (dot-dashed line), showing the asymptotic regime for large z. We use $b^2=2/3$. We also plot the ratio $M_4^2/M_*^3$ (lower panel) between 4-dimensional coupling and fundamental 5-dimensional Planck scale as a function of $\alpha$.} 
\end{figure}

To investigate this possibility, let us look at the asymptotic behavior of $U(z)$. In Fig.~4 (upper panel) we plot $z^2U(z)$, indicating that we have $U(z)\sim1/z^2$ for large $z$, even for large values of the parameter $\alpha$. We see from this figure that the asymptotic regime is better achieved for smaller values of $\alpha$, whereas for larger values one must consider still larger values of $z$. For instance, for $-30<z<30$, Fig.~4 shows that we can study analytically the correction for the Newtonian potential in the range $0<\alpha<0.1$. This is the region where our analytic expansion for small $\alpha$ agrees sensibly with the numerical simulations. We find $U(z)\sim(15/4)/z^2+(45/16)\pi a\alpha/z^3+{\cal O}(1/z^5)$ for large $z$. Note that the leading term of the expansion does not depend on $\alpha$. It is well known that potentials which asymptotes as $U(z)\sim\beta(\beta+1)/z^2$ gives a correction for the Newtonian potential ${\cal O}(1/R^{2\beta})$ for two massive objects at a distance $R$ from each other. In our case, we have $\beta=3/2$, and this gives the correction ${\cal O}(1/R^3)$ for the Newtonian potential, independent of $\alpha$. This is the same correction given by at the standard Randall-Sundrum scenario, and it confirms that the model localizes gravity for all chosen values of $\alpha$ where the expansion applies. The simulations also indicate that the same applies for larger values of $\alpha$. This means that a whole class of braneworld models were constructed with different properties related to matter distribution and geometry, all being able to localize gravity. 

Another point is that the next to leading term depends on $\alpha$ for the Schr\"odinger potential. It shows that for larger $\alpha$ the asymptotic regime is achieved for larger values of $z$, as demonstrated by the simulations shown in Fig.~4. In order to better see the influence of $\alpha$ in the gravitational interaction we remind that the Newtonian potential are corrected by the contribution of all massive modes, solutions of the Schr\"odinger-like equation, as 
\ben
\label{UR}
U(R)=G \frac1{R} + \frac 1{M_*^3}\int_0^{\infty}{dm \frac{e^{-mR}}R |\psi_m(0)|^2},
\een 
where the 4-dimensional coupling is $G=M_4^{-2}$, and $M_*$ is the fundamental 5-dimensional Planck scale, and the integration is considered at the brane position $z=0$. We can write an expression relating the two scales as
\ben
M_4^2=M_*^3\int_{-\infty}^{+\infty} dz e^{3A(z)}.
\label{eqMM}
\een
In this way one can see that those scales are related to the integral of a function depending on the warp factor. In Fig.~4 (lower panel) we plot the ratio $M_4^2/M_*^3$ between the two scales. Note the greater importance of $M_4$ for larger values of $\alpha$. Since $G=1/M_4^2$, this means that smaller values of $\alpha$ contribute to increase the intensity of the gravitational interaction, and that gravity localization is then favored. From Eq.~(\ref{typeIAz}) we can write $M_4^2/M_*^3=2/a+\alpha a^2/3$ for small $\alpha$. This is related to the fact that the Schr\"odinger potential achieves the asymptotic region at smaller values of $z$, for smaller values of $\alpha.$ This effect seems to be in perfect agreement with the former result, which has shown that the brane thickness decreases with decreasing values of $\alpha$, since in a thicker brane, gravity localizes less importantly.  

\subsection{Type II model}

We now study models with the function $F=-X^2/2$. In this case, the equation of motion of the scalar field is
\ben
\frac{3}{2}\phi^{\prime2} \phi^{\prime\prime} + 2\phi^{\prime3}A^\prime - V_\phi = 0
\een
Using the  first order formalism, the equation for $\phi^\prime$ is 
\ben\label{firstF2}
\phi^\prime = W_\phi^{\frac13}
\een
and the potential has the form
\ben
V(\phi)=\frac38 W_\phi^{\frac 43} - \frac13 W^2
\een

We choose an explicit function
\ben
W(\phi)=9 a^3 b^2 \sin(b\phi) (2+\cos^2(b\phi))
\een
Here the potential has the explicit form
\begin{eqnarray}
V(\phi)&=&\frac{243}8 a^4 b^4 \cos^4(b\phi) \nonumber \\
&&- 27 a^6 b^4 \sin^2(b\phi) \left(2+\cos^2(b\phi)\right)^2
\end{eqnarray}
In Fig.~5 we plot the potential for $a=1$ and several values of $b$, and for $b=1$ and several values of $a$. We note that in the first case, for $a=1$ the effect of increasing $b$ is to deepen and narrow the potential wells. In the second case, however, for $b=1$ and for $a$ increasing, each potential well deepens and widens, and this nicely contributes to distinguish the two cases, as we show below.    

\begin{figure}[htbp]
\centering
\includegraphics[width=6.0cm,height=5cm]{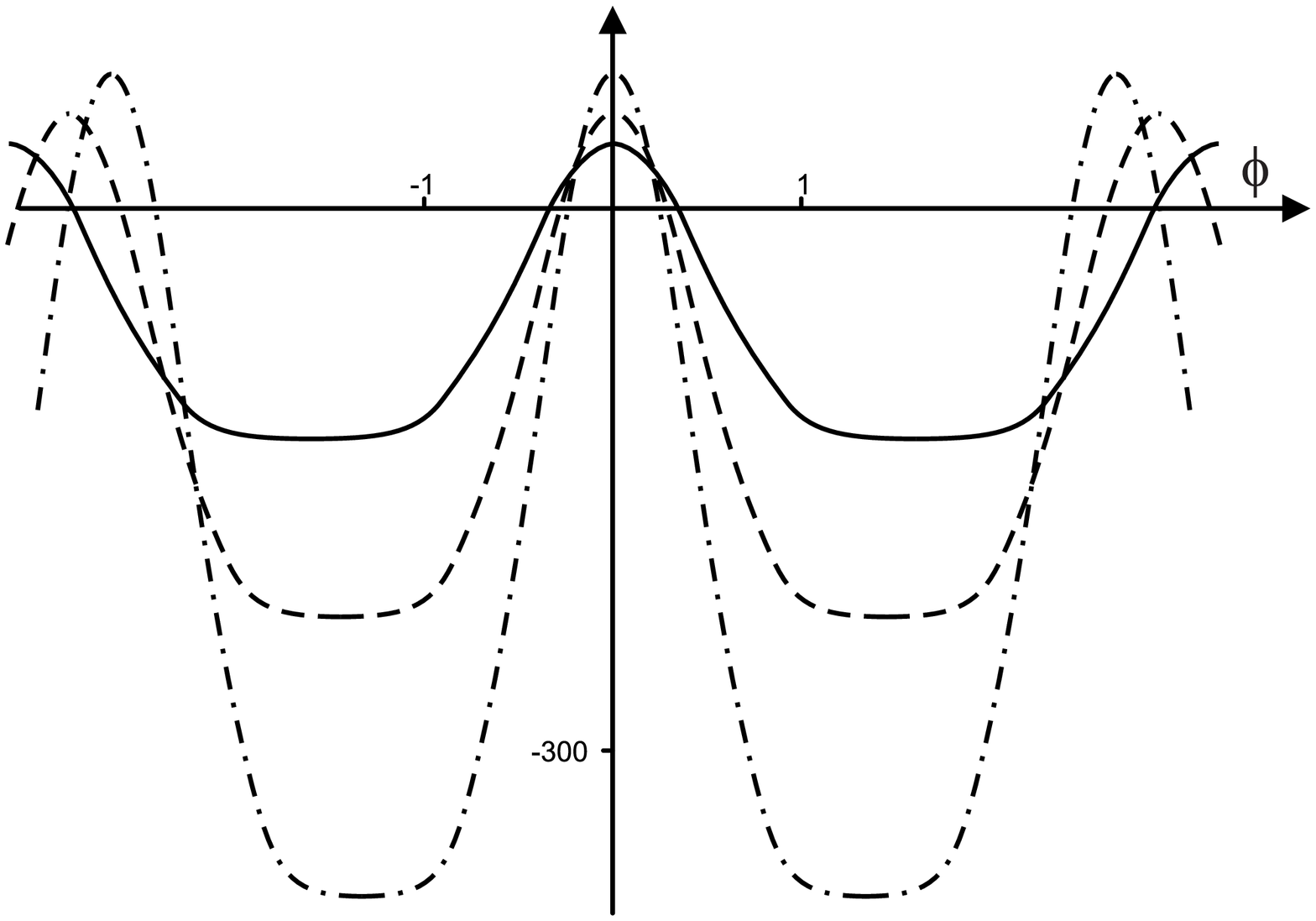}
\includegraphics[width=6.0cm,height=5cm]{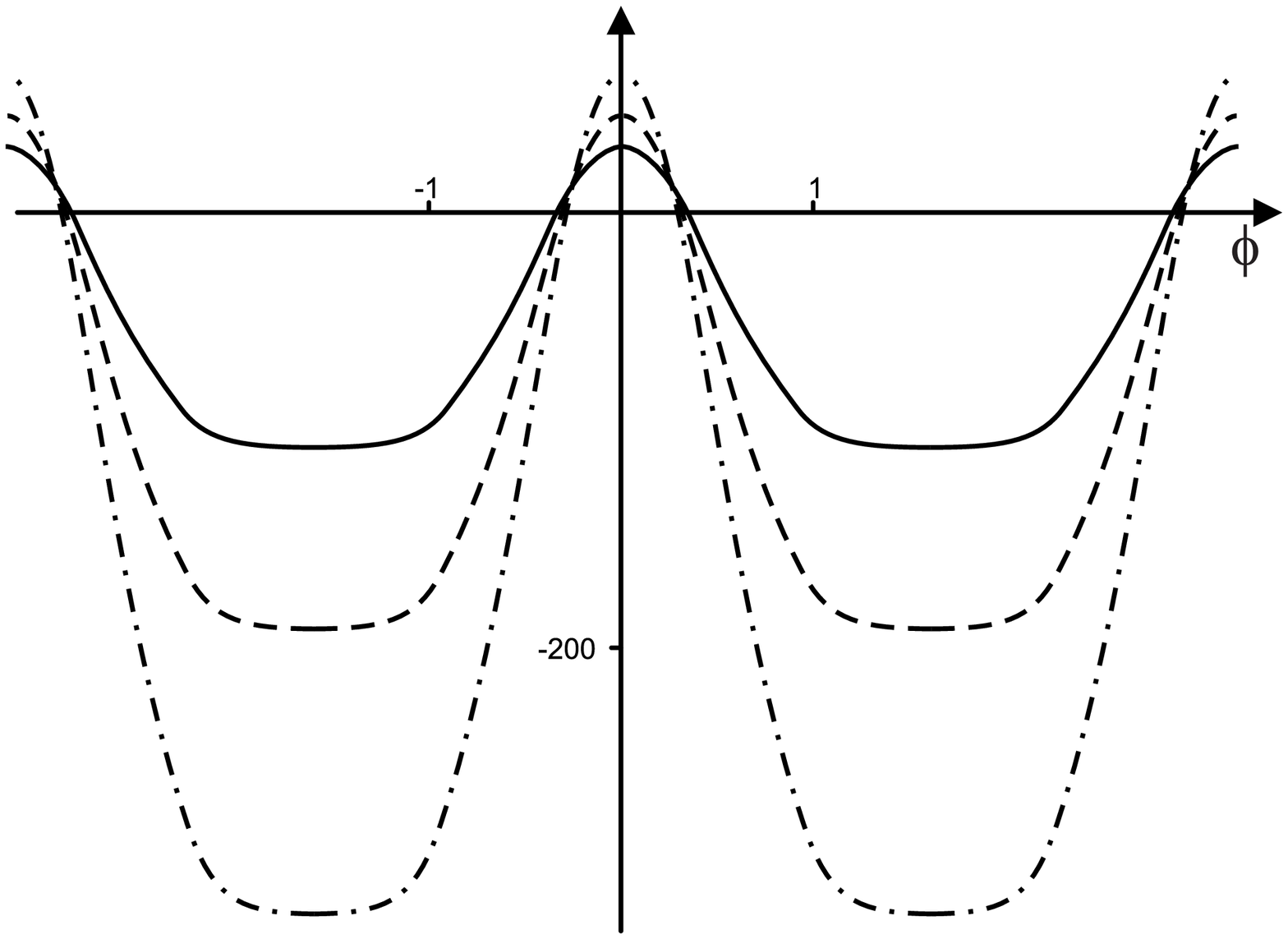}
\caption{Plots of the potential for $a=1$ (upper panel) and for $b=1$ (solid line), $b=1.1$ (dashed line) and $b=1.2$ (dot-dashed line), and for $b=1$ (lower panel), and for $a=1$ (solid line), $a=1.1$ (dashed line) and $a=1.2$ (dot-dashed line).}
\end{figure}

The scalar field obeys $\phi^\prime=3ab\cos(b\phi)$, with the solution  
\ben
\phi(y)=\frac1b \arcsin(\tanh(3ab^2y))
\een
We can also get for $A$ 
\ben
A(y)=-\frac16{a}^{2}\tanh^2\left( 3a{b}^{2}y \right)+\frac23{a}^{2}\ln \left({\sech}\left( 3\,a{b}^{2}y\right)\right) 
\een
which results in the warp factor depicted in Fig.~6, for $b=1$, with $a=1,1.1$, and $1.2$; we are not showing the other plots, for $a=1$ and for $b$ varying,
because they are essentially the same.

\begin{figure}[htbp]
\centering
\includegraphics[width=6.0cm,height=5cm]{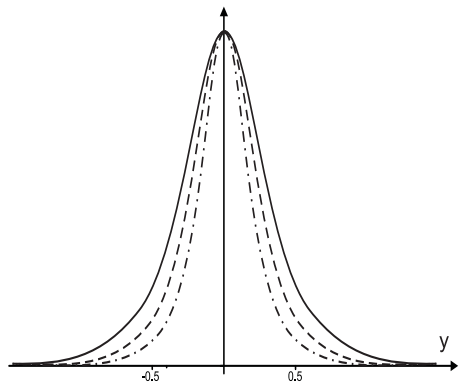}
\caption{Plots of the warp factor for $b=1$ and for $a=1$ (solid line), $a=1.1$ (dashed line) and $a=1.2$ (dot-dashed line).}
\end{figure}

\begin{figure}[htbp]
\centering
\includegraphics[width=6.0cm,height=5cm]{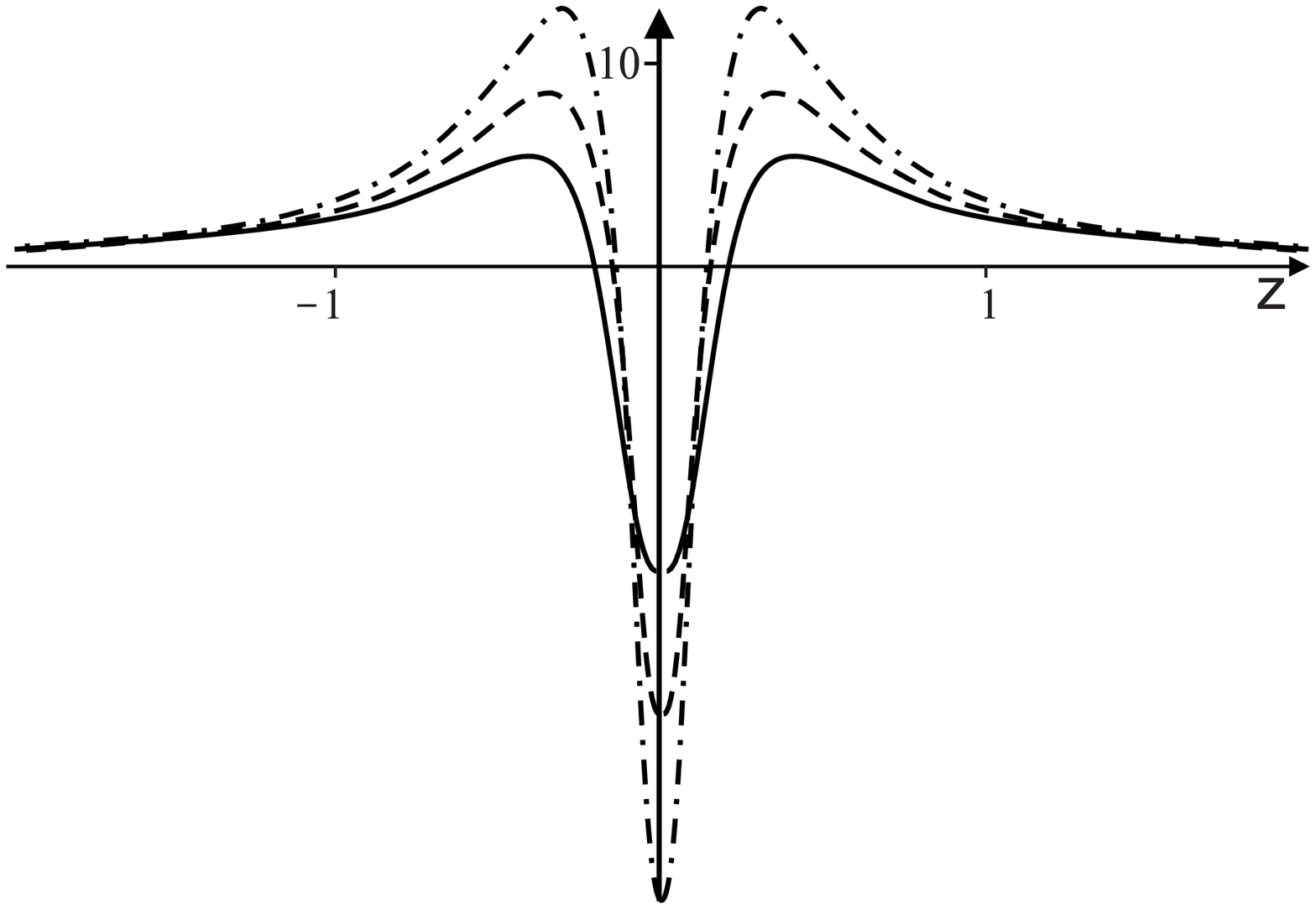}
\includegraphics[width=6.0cm,height=5cm]{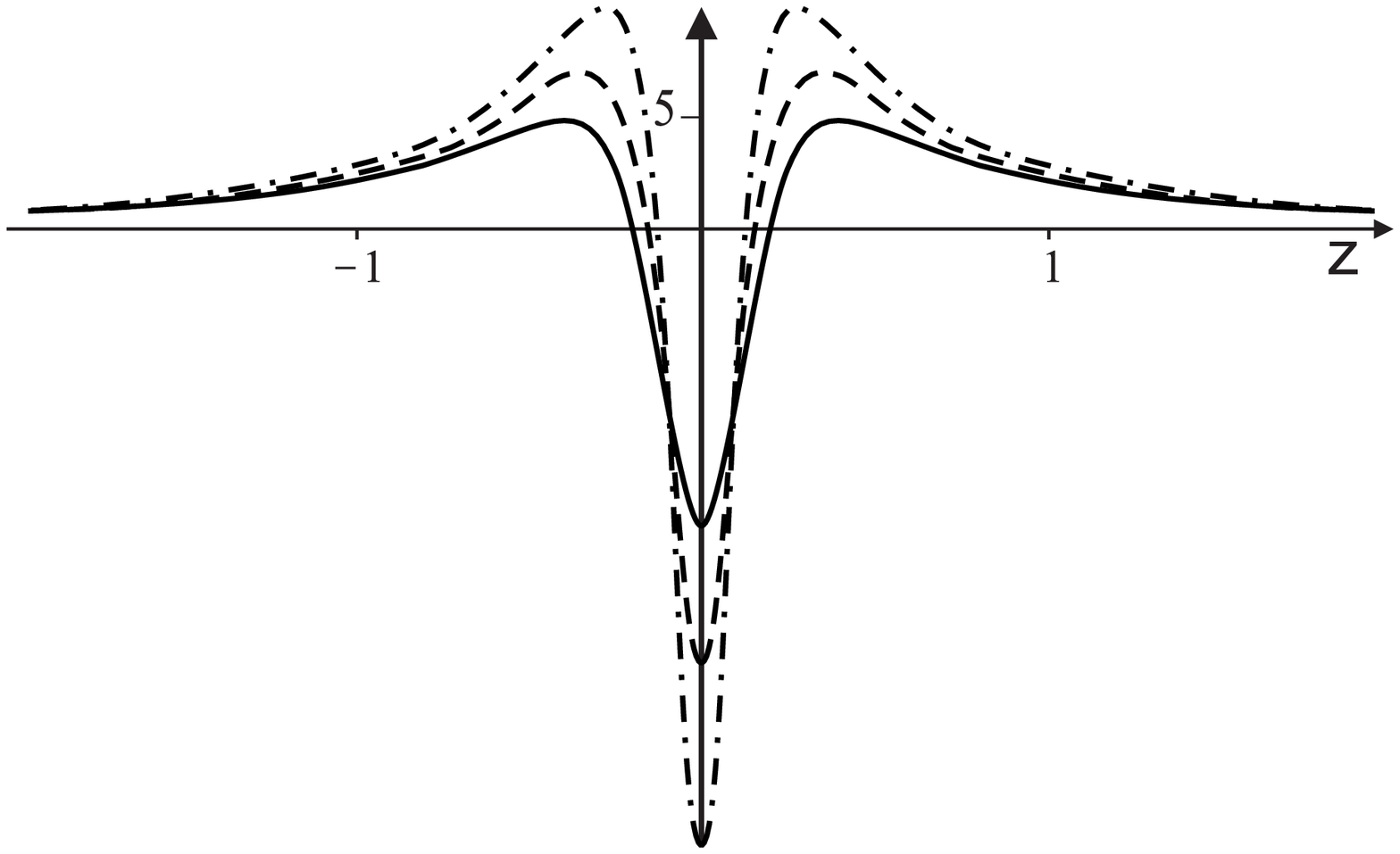}
\caption{Plots of the Schr\"odinger-like potential $U(z)$ for type II model. In the upper panel $a=1$ and 
$b=1$ (solid line), $b=1.1$ (dashed line) and $b=1.2$ (dot-dashed line). In the lower panel $b=1$ and   
$a=1$ (solid line), $a=1.1$ (dashed line) and $a=1.2$ (dot-dashed line). }\label{sch}
\end{figure}

\begin{figure}[htbp]
\centering
\includegraphics[width=6.0cm,height=5.0cm]{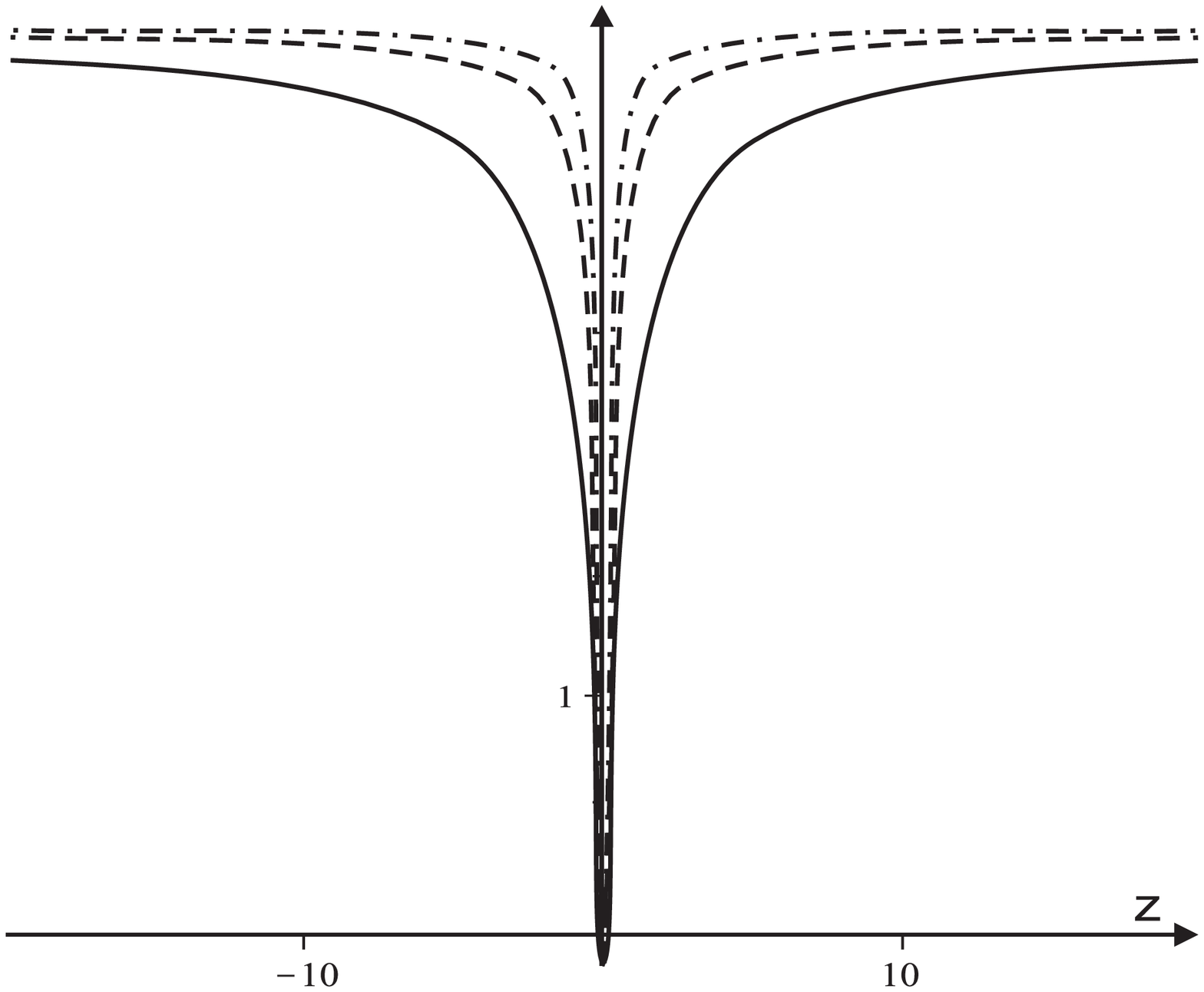}
\includegraphics[width=6.0cm,height=5.0cm]{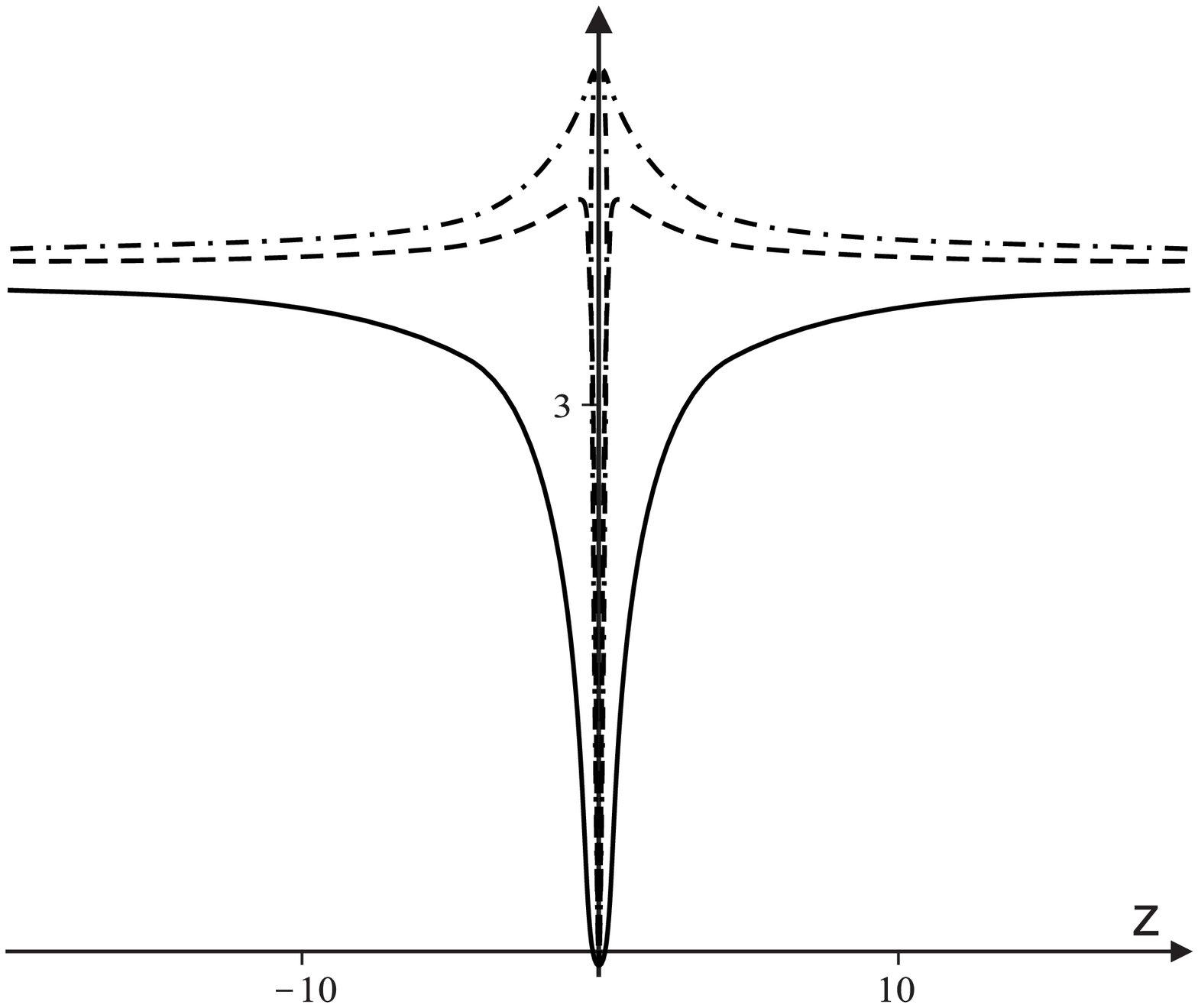}
\caption{Plots of $z^2U(z)$ for the type II model. In the upper panel $a=1$ and $b=1$ (solid line), $b=2$ (dashed line) and $b=3$ (dot-dashed line). In the lower panel $b=1$ and $a=1$ (solid line), $a=2$ (dashed line) and $a=3$ (dot-dashed line).}\label{zsq2}
\end{figure}

In the plots of Fig.~7, we see that gravity localization seems to be favored in the case $b=1$, for increasing $a$, since there we see that the height of the maxima of $U(z)$ is higher then in the other case, with $a=1$ for increasing $b$. However, to check this behavior quantitatively, we repeated the procedure included in Sec.~\ref{sec:typeI} in order to numerically obtain the functions $z(r),A(z)$ and $U(z)$.  
The function $U(z)$ was graphically analyzed to investigate the influence of the parameters $a$ and $b$ for gravity localization.
This is done in Fig. \ref{sch}, in which one displays $U(z)$ for several values of parameters $a$ and $b$, and there we note 
that: i) both figures have similar volcano profile. However, the numerical analysis shows that fixing $b=1$ with $1<a<3$ leads to larger extremes for the potential than fixing $a=1$ with $1<b<3$. In this way, we can infer that the parameter $a$ has greater influence for gravity localization; ii) fixing one of the parameters $a$ or $b$, the increasing of the second parameter leads to a corresponding increasing on the maxima and minima of the potential, indicating that
the increasing the parameters favors gravity localization. This is better seen in Fig. \ref{zsq2}, where we study $z^2U(z)$.  

In Fig. \ref{zsq2} one notes that the behavior of $U(z)$ suggests that $U(z)\sim1/z^2$ for $z>>1$. Writing $U(z)=\beta(\beta+1)/z^2$ we can, as in the analysis of the type I model, determine $\beta$ such that the Newtonian potential has correction of order $1/R^{2\beta}$ for large separation $R$ between unit masses. Now from these figures we can conclude that: i) for $a=1$ and fixed $z_{max}$, the asymptotic approximation is easier obtained for larger values of $b$. On the other hand, larger values of $b$ leads to larger values of $\beta$, and this corresponds to a correction with a greater power law for the Newtonian potential. This agrees with our previous analysis that larger values of $a$ or $b$ favor gravity localization; ii) for $z_{max}=20$ and $a=1$, we can affirm that the asymptotic region where $U(z)=(\beta)(\beta+1)/z^2$ is achieved with considerable precision for $b\ge2$. From the numerical values obtained we can estimate that:
for $b=2$, we get $(\beta)(\beta+1)=3.7180$, $\beta=1.492$, and a correction  $\mathcal{O}(1/R^{2.98})$ for the Newtonian potential (compare with $\mathcal{O}(1/R^{3})$ for Randall-Sundrum model); for $b=3$, we get $(\beta)(\beta+1)=3.7338$, $\beta=1.496$, and a correction  $\mathcal{O}(1/R^{2.99})$ for the Newtonian potential.

We further note that for $z_{max}=20$ and $b=1$ we can affirm that the asymptotic region is achieved for $a\ge2$. With the same procedure we estimate that:
for $a=2$, we get $(\beta)(\beta+1)=3.7809$, $\beta=1.508$, and a correction  $\mathcal{O}(1/R^{3.02})$ for the Newtonian potential; for $a=3$ we get $(\beta)(\beta+1)=3.8565$, $\beta=1.526$, and a correction  $\mathcal{O}(1/R^{3.05})$ for the Newtonian potential.

The above comments agree with the previous observation that larger values for $a$ and $b$ lead to corrections for the Newtonian potential with larger power law. The corrections are near to $\mathcal{O}(1/R^{3})$ obtained from Randall-Sundrum model for small parameters, and tend to be larger for larger $a$ and $b$, showing that gravity is then easier localized.

If we compare the case $a=1,b=3$ (correction $\mathcal{O}(1/R^{2.99})$) with the case $a=3,b=1$ (correction $\mathcal{O}(1/R^{3.05})$), we see that the increasing of the parameter $a$ is more effective for gravity localization in comparison with similar increasing of parameter $b$.

\section{Ending comments}
\label{end}

In this work we have investigated generalized braneworld models, that is, models in which the standard gravity action is extended to include
scalar fields with generalized dynamics, with the Lagrange density having the nonstandard form ${\cal L}(\phi,X)=F(X)-V(\phi)$. This study is a continuation 
of our program to investigate the scalar field behavior under the presence of generalized dynamics \cite{Bazeia:2007df,blm}. In particular, in the present study we have included two distinct families of models, one given by $F(X)=X-\alpha X |X|$, and the other by $F(X)=-X^2$. 

An interesting and general result is that the proposed scenario, in which gravity acts standardly and the scalar field has generalized dynamics, is linearly stable, capable of localizing gravity in a way similar to the standard case. 

Other investigations included in this work engender both analytical and numerical results. For the type I model, with $F(X)=X-\alpha X |X|,$ we have presented analytical results up to first order in $\alpha$ for $\alpha$ very small, and our numerical study confirms the correctness of the perturbative expansion up to first order in $\alpha$. In this case, for $\alpha$ small, we have calculated analytically the corrections to the Newtonian potential, showing the localization of gravity. The ratio between the masses leads to a nice estimate of how $\alpha$ quantitatively affects the gravitational interaction. And numerically, we could extend this result to much larger values of $\alpha$. As another interesting result, we have shown that gravity localization is more effective at smaller values of $\alpha$. It seems that for $F(X)=X+\alpha|X|X$, the robustness of the model weakens for increasing values of $\alpha$, as we get away from the standard braneworld scenario.

The type II model is more involved. However, thanks to the first-order framework put forward in \cite{blm}, we could study it and obtain analytic solutions for both the scalar field and warp factor. The analytic solutions has helped us to ease the subsequent numerical investigations, to study gravity localization. The results show that, like in the former case, this new and well distinct family of models engenders similar behavior, and it is also capable of localizing gravity. For the two families of models, we can also control gravity localization with the specific form of the potential $V(\phi)$, an effect that also appears in the braneworld model with the scalar field with standard dynamics.

There are other possibilities of study. For instance, in the case of the type II model, we can find compacton solutions for the scalar field, so we can also investigate this case, in a way similar to the study done in \cite{Adam}. Another possibility is to extend the present investigations to the case of bent brane, as we have already done in some of our work in \cite{first2} in the case of standard dynamics. We can also consider the harder case, in which we also change $R$ to $F(R)$, extending the $F(R)$ brane study done in \cite{abmp} to this new $F(R,X)$ braneworld scenario. We hope to report on these issues in another work in the near future.  

The authors would like to thank CAPES, CNPq, CNPq-MCT-CT-Energ and PRONEX-CNPq-FAPESQ for partial support.



\begin{thebibliography}{99}
\bibitem{RS}
L.~Randall and R.~Sundrum, Phys.\ Rev.\ Lett.\  {\bf 83}, 4690 (1999) [arXiv:hep-th/9906064].
\bibitem{GW}
W.~D.~Goldberger and M.~B.~Wise, Phys.\ Rev.\ Lett.\  {\bf 83}, 4922 (1999) [arXiv:hep-ph/9907447].
\bibitem{first1}
M. Cvetic and H.H. Soleng, Phys. Rev. D 51, 5768 (1995) [arXiv:hep-th/9411170]; Phys. Rep. 282, 159
(1997) [arXiv:hep-th/9604090]; K. Skenderis and P.K. Townsend, Phys. Lett B {\bf468}, 46 (1999) [arXiv:hep-th/9909070];
O. DeWolfe, D.Z. Freedman, S.S. Gubser and A. Karch, Phys. Rev. D {\bf62}, 046008 (2000) [arXiv:hep-th/9909134];
C. Csaki, J. Erlich, T. Hollowood and Y. Shirman, Nucl. Phys. B {\bf581}, 309 (2000) [arXiv:hep-th/0001033]; C. Csaki, J. Erlich,
C. Grojean, and T. Hollowood, Nucl. Phys. B {\bf 584}, 359 (2000) [arXiv:hep-th/0004133]; 
M.Gremm, Phys. Lett. B {\bf478}, 434 (2000) [arXiv:hep-th/9912060]; M. Porrati, Phys. Lett. B {\bf498}, 92 (2001) [arXiv:hep-th/0011152]; F.A. Brito, M. Cvetic, S.-C. Yoon, Phys. Rev. D64 (2001) 064021; [arXiv:hep-ph/0105010]; M. Cvetic, N.D. Lambert, Phys. Lett. B {\bf540}, 301 (2002) [arXiv:hep-th/0205247]; A. Melfo, N. Pantoja, A. Skirzewski, Phys. Rev. D 67, 105003 (2003)[arXiv:gr-qc/0211081]; D. Bazeia, F.A. Brito, and J.R. Nascimento, Phys. Rev. D {\bf68}, 085007 (2003) [arXiv:hep-th/0306284]; D. Bazeia, C. Furtado and A.R. Gomes, JCAP 0402, 002 (2004) [arXiv:hep-th/0308034]; D. Bazeia and A.R. Gomes, JHEP {\bf0405}, 012 (2004) [arXiv:hep-th/0403141]; O. Castillo-Felisola, A. Melfo, N. Pantoja, A. Ramirez, Phys. Rev. D 70, 104029 (2004) [arXiv:hep-th/0404083]; K. Takahashi, T. Shiromizu, Phys. Rev. D 70, 103507 (2004); R. Guerrero, R. Omar Rodrigues, and R. Torrealba, Phys. Rev. D 72, 124012 (2005) [arXiv:hep-th/0510023].
\bibitem{first2}
D.Z. Freedman, C. Nunez, M. Schnabl and K. Skenderis, Phys. Rev. D {\bf69}, 104027 (2004) [hep-th/0312055]; A. Celi, A. Ceresole, G. Dall'Agata, A. Van Proeyen and M. Zagermann, Phys. Rev. D {\bf71}, 045009 (2005) [hep-th/0410126]; M. Zagermann, Phys. Rev. D {\bf71}, 125007 (2005) [hep-th/0412081];
D. Bazeia, C.B. Gomes, L. Losano and R. Menezes, Phys. Lett. B {\bf633}, 415 (2006) [astro-ph/0512197]; V.I. Afonso, D. Bazeia, L. Losano, Phys. Lett. B {\bf634}, 526 (2006) [hep-th/0601069]; K. Skenderis and P. K. Townsend, Phys. Rev. Lett. {\bf96}, 191301 (2006) [hep-th/0602260]; D. Bazeia, F.A. Brito, L. Losano, JHEP {\bf0611}, 064 (2006) [arXiv:hep-th/0610233]; K. Skenderis and P. K. Townsend, J. Phys. A {\bf40}, 6733 (2007) [hep-th/0610253].
\bibitem{OT}
A. Ceresole, G. Dall'Agata, JHEP {\bf0703}, 110 (2007) [arXiv:hep-th/0702088]; W. Chemissany, A. Ploegh, T. Van Riet, Class. Quant. Grav. {\bf24}, 4679 (2007) 
[arXiv:0704.1653]; E.A. Bergshoeff, J. Hartong, A. Ploegh, J. Rosseel, and D. Van den Bleeken, JHEP {\bf0707}, 067 (2007)b [arXiv:0704.3559]; L. Cardoso, A. Ceresole, G. Dall'Agata, J.M. Oberreuter, and J. Perz, JHEP {\bf0710}, 063 (2007) [arXiv:0706.3373];
B. Janssen, P. Smyth, T.  Van Riet, B. Vercnocke, JHEP {\bf0804} 007 (2008) [arXiv:0712.2808]; M. Cvetic and M. Robnik, [arXiv:0801.0801]; S. Ferrara, A. Gnecchi, and A. Marrani [arXiv:0806.3196]; Y.-X. Liu, L.-D. Zhang, L.-J. Zhang, and Y.-S. Duan, [arXiv:0804.4553]. 
\bibitem{AP}
C.~Armendariz-Picon, T.~Damour and V.~F.~Mukhanov, Phys.\ Lett.\  B {\bf 458}, 209 (1999), [arXiv:hep-th/9904075];
T.~Chiba, T.~Okabe and M.~Yamaguchi, Phys.\ Rev.\  D {\bf 62}, 023511 (2000), [arXiv:astro-ph/9912463]; 
C.~Armendariz-Picon, V.~F.~Mukhanov and P.~J.~Steinhardt, Phys.\ Rev.\ Lett.\  {\bf 85,} 4438 (2000), [arXiv:astro-ph/0004134].
\bibitem{Babichev:2006cy}
E.~Babichev, Phys.\ Rev.\  D {\bf 74}, 085004 (2006), [arXiv:hep-th/0608071].
\bibitem{Bazeia:2007df}
D.~Bazeia, L.~Losano, R.~Menezes and J.C.R.~Oliveira, Eur.\ Phys.\ J.\  C {\bf 51}, 953 (2007) [arXiv:hep-th/0702052].
\bibitem{Adam:2007ij}
C.~Adam, J.~Sanchez-Guillen and A.~Wereszczynski, J. Phys. A {\bf40}, 13625 (2007) [arXiv:0705.3554].
\bibitem{Babichev:2007tn}
E.~Babichev, [arXiv:0711.0376].
\bibitem{Olechowski:2008bh}
M.~Olechowski, [arXiv:0801.1605].
\bibitem{Adam}
C. Adam, N. Grandi, J. Sanchez-Guillen, and A. Wereszczynski, J. Phys. A {\bf41}, 212004 (2008) [arXiv:0711.3550];
C. Adam, N. Grandi, P. Klimas, J. Sánchez-Guillén, and A. Wereszczynski, [arXiv:0805.3278].  
\bibitem{blm}
D. Bazeia, L. Losano and R. Menezes, Phys. Lett. B {\bf668}, 246 (2008), [arXiv:0807.0213].
\bibitem{abmp}
V.I. Afonso, D. Bazeia, R. Menezes, and A.Yu. Petrov, Phys. Lett. B {\bf658}, 71 (2007), [arXiv:0710.3790].   
\end{thebibliography}
\end{document}